\documentclass{article}
\usepackage{dcase2021_techrep,amsmath,graphicx,url,times,booktabs, tabularx}
\usepackage{enumitem}
\usepackage{float}
\usepackage{multirow}
\usepackage{booktabs}
\usepackage[table,xcdraw]{xcolor}
\usepackage{soul}

\title{Sound Event Localization and Detection using Squeeze-Excitation Residual CNNs}

\name{Javier Naranjo-Alcazar$^{1,2}$,
      Sergi Perez-Castanos$^{2}$,
      Maximo Cobos$^{2}$,
      Francesc J. Ferri$^{2}$,
      Pedro Zuccarello$^{1}$
      }

\address{$^1$ Instituto Tecnológico de Informática, València, Spain \{jnarnajo, pzuccarello\}@iti.es\\         
$^2$ Universitat de Val\`encia, Burjassot, Spain, \{pecaser@alumni.uv.es, \{maximo.cobos, francesc.ferri\}@uv.es\}\\
 }
 
\begin{document}

\ninept
\maketitle

\begin{sloppy}

\begin{abstract}
Sound event localisation and detection (SELD) is a problem in the field of automatic listening that aims at the temporal detection and localisation (direction of arrival estimation) of sound events within an audio clip, usually of long duration. Due to the amount of data present in the datasets related to this problem, solutions based on deep learning have positioned themselves at the top of the state of the art. Most solutions are based on 2D representations of the audio (different spectrograms) that are processed by a convolutional-recurrent network. The motivation of this submission is to study the squeeze-excitation technique in the convolutional part of the network and how it improves the performance of the system. This study is based on the one carried out by the same team last year. This year, it has been decided to study how this technique improves each of the datasets (last year only the MIC dataset was studied). This modification shows an improvement in the performance of the system compared to the baseline using MIC dataset.
\end{abstract}


\section{Introduction}
\label{sec:intro}

\par Sound event localisation and detection (SELD) corresponds to the machine listening problem that aims to detect and localise a sound event in an audio clip of typically long duration \cite{Kapka2019, Cao2019, Xue2019}. The detection consists in correctly classifying the sound event into one of the predefined classes while setting the time at which the event starts and ends. For several editions of DCASE, this task only attempted to solve the SED problem. On the other hand, localisation consists in estimating the direction of arrival (DOA) of the source producing the event in terms of azimuth and elevation angles. SELD proposes a joint problem involving these two areas, for which a single system capable of performing both detection and localisation must be proposed. For an intelligent system to be able to estimate directional information, audio signals must have been recorded by a set of microphones (multi-channel audio input).

\par The task of Sound Event Location and Detection (SELD) has been constantly in evolution in the scope of the DCASE Challenge until reaching the problem presented in this edition. In 2013 \cite{Stowell2015}, the problem to be solved was known as Sound Event Detection (SED). In the 2016 \cite{Mesaros2018_TASLP} and 2017 \cite{Mesaros2019_TASLP} editions, this problem was again proposed as a task. In this case, the objective was to create a system capable of detecting the onset and offset of sound events while correctly classifying to which class these events belong. The first time that event localisation was raised in addition to the SED problem was in the 2019 edition \cite{Adavanne2018_JSTSP, Adavanne2019_DCASE}. Last year, 2020 \cite{politis2020dataset}, the dataset was modified in addition to the metrics (setting a threshold of 20° in the detection metrics). This edition incorporates a new consideration which is the existence of directional interferences, meaning sound events out of the target classes that are also point-like in nature. This is a much more accurate recreation of a real environment \cite{politis2020overview}.

\begin{figure*}[t]
  \centering
  \centerline{\includegraphics[scale=0.55]{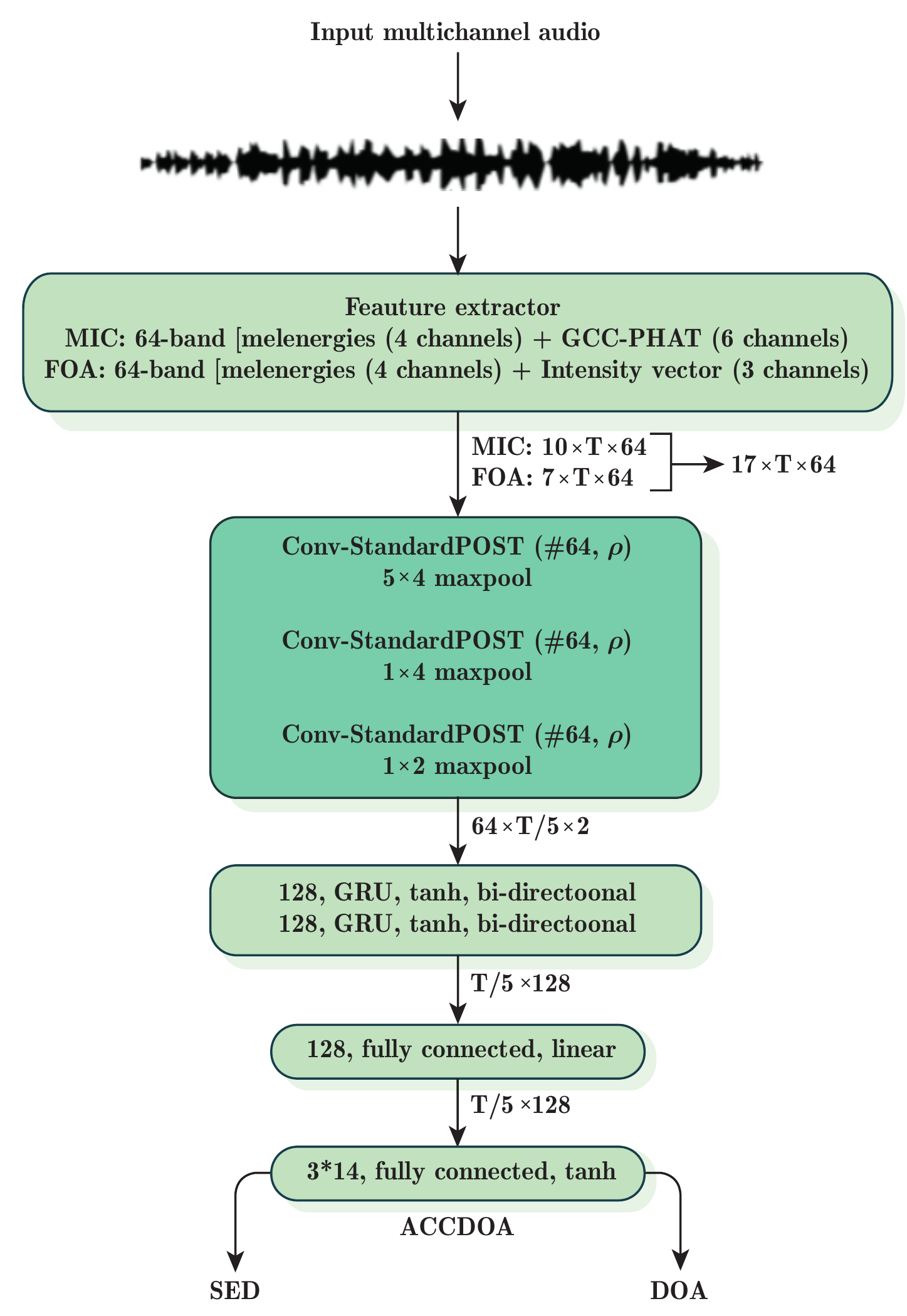}}
  \caption{SELD framework proposed in this work. The most highlighted block corresponds to the change made in this task. The lighter blocks have the same configuration as in the baseline. $\rho$ indicates the ratio parameter. In this work, $\rho=1$.}
  \label{fig:seld_structure}
\end{figure*}

\par The work done in this edition can be seen as a extension of the work done in the last edition \cite{Naranjo-Alcazar2020_task3_report}. Last year, an analysis was made of how squeeze-excitation and residual techniques \cite{roy2018concurrent, naranjoalcazar2020acoustic} (applied in the convolutional part of the system) can lead to a more robust system without any extra modification of the baseline. For this, the \emph{Conv-StandardPOST} block was implemented and it was analysed how the different ratios ($\rho$) influenced the system. Furthermore, it was compared with a residual block that did not implement any squeeze-excitation technique, please see Fig.~\ref{fig:seld_residual}. According to the results obtained in the Challenge, the block with $\rho=1$ obtained the best position. So, since this study was performed only using the MIC dataset, this year we will analyse how the \emph{Conv-StandardPOST} block with a fixed ratio behaves with each of the datasets.

\par This technical report is organized as follows: Section \ref{sec:method} introduces the network presented as the baseline and the modification done in this work to achieve the improvement. Section~\ref{sec:exp_details} explains the dataset used and the training procedure. Section \ref{sec:results} shows the results obtained by the framework implemented and Section \ref{sec:conclusion} concludes our work.

% Below is an example of how to insert images. 
% -------------------------------------------------------------------------

% \begin{equation}
%   \label{eqn:wave_equation}
%     \Delta^2p(x,y,z,t)-
%     \displaystyle\frac{1}{c^2}\frac{\partial^2p(x,y,z,t)}{\partial t^2}=0,
% \end{equation}

\section{Method}\label{sec:method}

\subsection{Baseline System}\label{subsec:baseline}

\par The baseline network is known as SELDnet \cite{Adavanne2018_JSTSP}. The main modification this year has been the elimination of the SED classification branch, adopting a joint training (ACCDOA) that unites SED loss and DOA in a homogenous regression vector loss \cite{Shimada2021}.

\par The first module of the system is the feature extractor. It obtains multi-channel audio representations. With the MIC dataset, 10 channels (GCC) are obtained and with the first-order ambisonics (FOA), 7 channels (Intensity vector). These representations were introduced in \cite{Cao2019}.

%\par The input to the network is 10-channel representation \textcolor{orange}{in MIC format of the dataset provided by the organizers of Task 3 (see Section~\ref{subsec:dataset} for an explanation about MIC format and other details of the dataset)} (\st{more insight of the dataset is provided in Section~\ref{subsec:dataset}}). The dimension of each channel is $T \times F$, where $T$ corresponds to the number of temporary bins and $F$ to the number of frequency bins. In this case, $F$ is set to 64 and $T$ corresponds to 300 temporal bins. Four of these time-frequency representations correspond to the log-Mel Spectrogram of each of the microphone signals, and the other six are the generalized cross-correlations (GCCs) between the microphone signals \cite{Cao2019}. The implementation of the baseline network can be found in this link\footnote{https://github.com/sharathadavanne/seld-dcase2020}.

\begin{figure}[t]
  \centering
  \centerline{\includegraphics[scale=0.85]{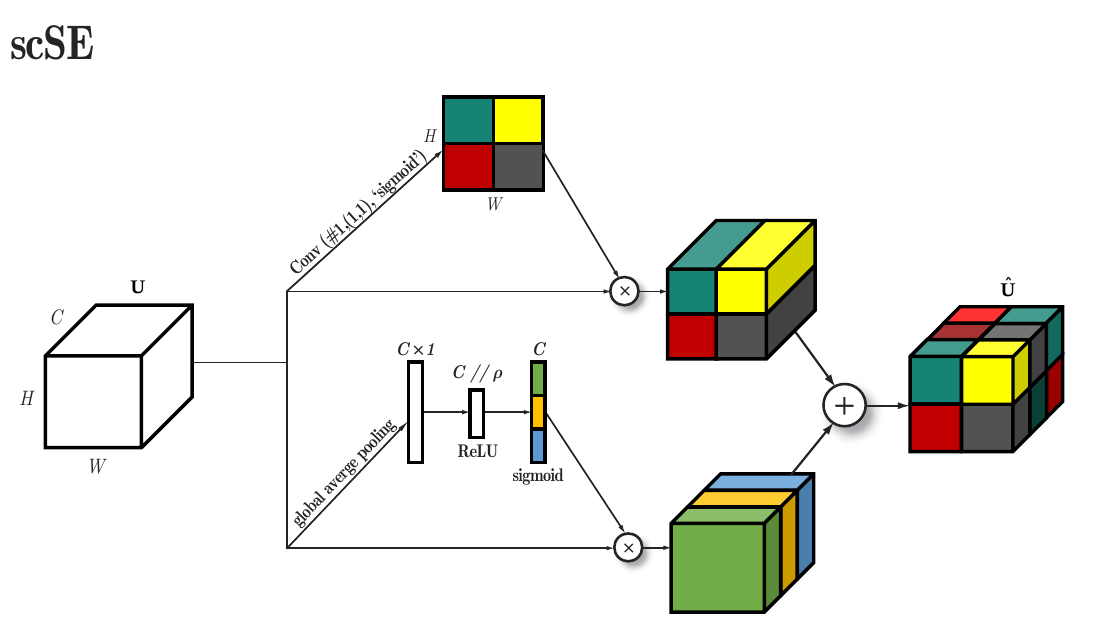}}
  \caption{scSE composed by Spatial Squeeze-Excitation (sSE) module (top branch) and channel Squeeze-Excitation (cSE) module (lower branch) \cite{naranjoalcazar2020acoustic, roy2018concurrent}} 
  \label{fig:scSE}
\end{figure}

\begin{figure}[t]
  \centering
  \centerline{\includegraphics[scale=0.45]{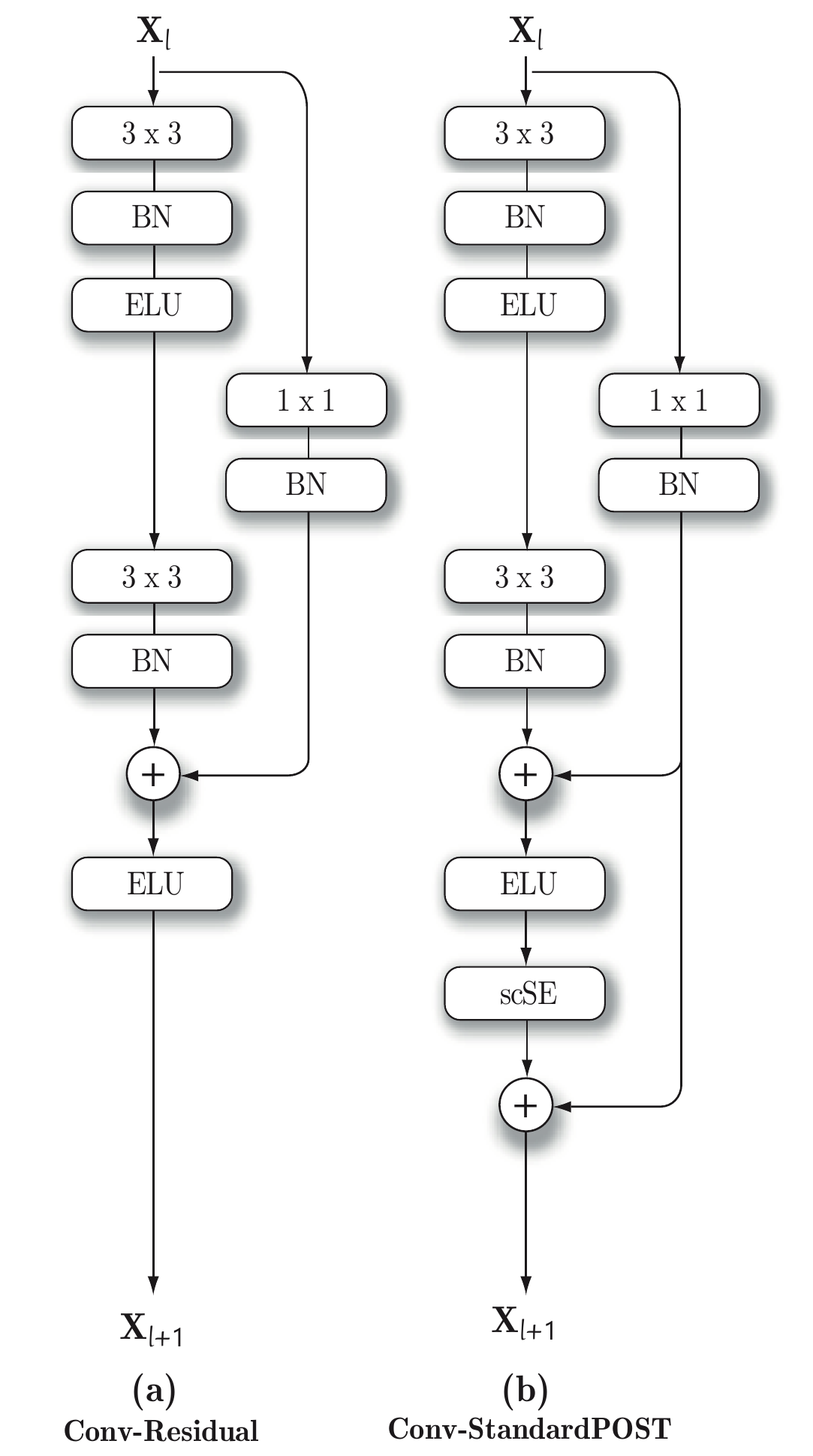}}
  \caption{Residual blocks analyzed in this paper. BN stands for Batch Normalization and scSE for squeeze-excitation module. Convolutional layers are indicated with the kernel size.}
  \label{fig:seld_residual}
\end{figure}

\subsection{Squeeze-Excitation Residual blocks and modifications to the baseline network}\label{subsec:SE}

\par This submission is understood as a extension of the work done in  \cite{naranjoalcazar2020acoustic, Naranjo-Alcazar2020_task3_report} where different different squeeze-excitation modules where studied \cite{hu2018squeeze} plus the contribution of two novel blocks using the \textit{Concurrent Spatial and Channel Squeeze and Channel Excitation} (scSE) configuration presented in \cite{roy2018concurrent}. The implementation of the scSE block is detailed in Figure~\ref{fig:scSE}. Following the conclusions of \cite{naranjoalcazar2020acoustic} and \cite{Naranjo-Alcazar2020_task3_report}, in the present work, the convolutional layers of SELDnet are replaced by the \textit{Conv-StandardPOST} blocks with $\rho=1$. The number of filters remains the same (64 filters).  The framework proposed in this work is shown in Figure~\ref{fig:seld_structure}.
% \st{In  \cite{naranjoalcazar2020acoustic}, the novel configuration labelled as \textit{Conv-StandardPOST} (see Figure~\ref{fig:seld_residual}(b)) showed the best results in treated ASC problem.}

\par The code used for this experimentation can be found in the  following link\footnote{https://github.com/Machine-Listeners-Valencia/seld-dcase2021}.

\section{Experimental details}\label{sec:exp_details}

\subsection{Dataset}\label{subsec:dataset}

\par The dataset used in this edition is the one defined as TAU-NIGENS Spatial Sound Events 2021. The major change this year is the addition of sound events that do not belong to the target classes. For a more detailed description of the dataset, visit the following link\footnote{http://dcase.community/challenge2021/task-sound-event-localization-and-detection} and the paper \cite{politis2020dataset}.

\par Concerning the usage of the samples (see Table~\ref{tab:kfold}), in the development phase, three folders are used for training, one for validation and one for testing. In this stage, the ground truth of all the samples is available. However, in the evaluation stage, 4 folders are used for training, 1 for validation and 2 for testing. In this case, the ground truth of the test samples is not available (Challenge results).

%\textcolor{blue}{The dataset is distributed differently depending to the stage. This means that in the development phase, three folders are used for training, one for validation and one for testing. In this stage, the ground truth of all the samples is available. In the evaluation stage, 4 folders are used for training, 1 for validation and 2 for testing. In this case, the ground truth of the test samples is not available, therefore, the results of the validation folder will be reported in this stage. The distribution can be seen in Table~\ref{tab:kfold}.}

\begin{table}[]
\centering
\begin{tabular}{cccc}
\toprule
Stage       & training & validation & test \\ \toprule
Development & 3-6      & 2          & 1    \\ \midrule
Evaluation  & 2-6      & 1          & 7-8 \\ \bottomrule
\end{tabular}
\caption{Distribution of the folders in the two stages. Each folder contains 100 samples.}
\label{tab:kfold}
\end{table}

\subsection{Training procedure}\label{subsec:training}

\par The training process is the same as that proposed in the baseline \cite{Adavanne2018_JSTSP, politis2020dataset}. However, it has been decided to implement a learning rate decay system. Thus, if the performance of the system is not improved within 15 epochs, the learning rate decreases by a factor of $0.5$. The training is terminated if there is no improvement in 30 epochs.

\section{Results}\label{sec:results}

\par In order to study the squeeze-excitation residual blocks in both datasets, it was decided to carry out the experimentation with $\rho = 1$ as it showed the best performance in last year submission \cite{Naranjo-Alcazar2020_task3_report}.

\subsection{Metrics}\label{subsec:metrics}

\par The metrics used in this task are known as location-dependent. The detection of an event will be considered correct if the angle prediction is below a threshold set at 20°.

%The metrics for detection make use of the error rate (ER\textsubscript{20º}) and the F-Score (F\textsubscript{20º}),

\par In this task, there are 2 metrics per output. Two metrics to measure the robustness of the detection are the error rate (ER\textsubscript{20º}) and the F-score (F\textsubscript{20º}) with a threshold of 20º. On the other hand, the localisation accuracy is measured with the localisation error (LE\textsubscript{CD}) and the localisation recall (LR\textsubscript{CD}). For more insight about metrics, please see \cite{politis2020overview}.

\subsection{Development stage}\label{subsec:dev}

The results obtained in both datasets independently and combined (concatenating the representations obtained with each one) are shown below (see Table~\ref{tab:results}). If the FOA dataset is considered, it can be seen that the metric that is improved is LR\textsubscript{CD}. However, LE\textsubscript{CD} decreases by 4º. On the other hand, the MIC dataset is greatly improved, the system shows a better performance in all metrics. The most improved metric is F\textsubscript{20º}, improving by 5.2 percentage points. Finally, if both representations are concatenated, no considerable improvement can be seen with respect to the FOA dataset. 

\begin{table}[H]
\centering
\begin{tabular}{c|c|c|c|c|c }
\toprule
  \emph{Framework} & \emph{Dataset} & \emph{ER\textsubscript{20º}} & \emph{F\textsubscript{20º}} & \emph{LE\textsubscript{CD}} & \emph{LR\textsubscript{CD}}\\ 
 \midrule
\rowcolor{gray!25} Baseline & FOA & 	0.69 & 	33.9\% & 24.1º & 	43.9\% \\ 
 \midrule
 Proposed & FOA & 0.71 & 31.9\% & 27.6º & \textbf{46.6\%}\\
 \midrule
 \rowcolor{gray!25}  Baseline & MIC & 	0.74 & 	24.7\% & 30.9º & 	38.2\% \\ 
  \midrule
 Proposed  & MIC & \textbf{0.72} & \textbf{30.2\%} & \textbf{29.4º} & \textbf{42.5\%}\\
 \midrule
 \rowcolor{gray!10} Proposed  & FOA-MIC & 	0.71 & 	31.3\% & 27.9º & 	46.7\% \\
 \bottomrule
\end{tabular}
\caption{Accuracy (\%) results obtained compare with the proposed baseline}
\label{tab:results}
\end{table}

\section{Conclusion}\label{sec:conclusion}

The motivation for this work is the study of squeeze-excitation techniques for the improvement of SED/DOA systems. For this purpose, it has been decided to modify only the convolutional part of the system and to follow the conclusions obtained in last year's edition. Despite the interferences present in this edition, it can be observed that, using the MIC dataset, all the metrics are improved. However, we cannot observe the same behaviour using the FOA dataset. This suggests further study if this dataset is to be used.

\section{Acknowledgements}\label{sec:acknow}

\par The participation of Dr. Cobos  and Dr. Ferri is supported by ERDF and the Spanish Ministry of Science,
Innovation and Universities under Grant RTI2018-097045-B-C21, as well as grants AICO/2020/154 and AEST/2020/012 from Generalitat Valenciana.

% -------------------------------------------------------------------------
% Either list references using the bibliography style file IEEEtran.bst
\bibliographystyle{IEEEtran}
\bibliography{refs}

\begin{thebibliography}{10}
\providecommand{\url}[1]{#1}
\def\UrlFont{\rmfamily}
\providecommand{\newblock}{\relax}
\providecommand{\bibinfo}[2]{#2}
\providecommand\BIBentrySTDinterwordspacing{\spaceskip=0pt\relax}
\providecommand\BIBentryALTinterwordstretchfactor{4}
\providecommand\BIBentryALTinterwordspacing{\spaceskip=\fontdimen2\font plus
\BIBentryALTinterwordstretchfactor\fontdimen3\font minus
  \fontdimen4\font\relax}
\providecommand\BIBforeignlanguage[2]{{%
\expandafter\ifx\csname l@#1\endcsname\relax
\typeout{** WARNING: IEEEtran.bst: No hyphenation pattern has been}%
\typeout{** loaded for the language `#1'. Using the pattern for}%
\typeout{** the default language instead.}%
\else
\language=\csname l@#1\endcsname
\fi
#2}}

\bibitem{Kapka2019}
S.~Kapka and M.~Lewandowski, ``Sound source detection, localization and
  classification using consecutive ensemble of crnn models,'' DCASE2019
  Challenge, Tech. Rep., June 2019.

\bibitem{Cao2019}
Y.~Cao, T.~Iqbal, Q.~Kong, M.~Galindo, W.~Wang, and M.~Plumbley, ``Two-stage
  sound event localization and detection using intensity vector and generalized
  cross-correlation,'' DCASE2019 Challenge, Tech. Rep., June 2019.

\bibitem{Xue2019}
W.~Xue, T.~Ying, Z.~Chao, and D.~Guohong, ``Multi-beam and multi-task learning
  for joint sound event detection and localization,'' DCASE2019 Challenge,
  Tech. Rep., June 2019.

\bibitem{Stowell2015}
D.~{Stowell}, D.~{Giannoulis}, E.~{Benetos}, M.~{Lagrange}, and M.~D.
  {Plumbley}, ``Detection and classification of acoustic scenes and events,''
  \emph{IEEE Transactions on Multimedia}, vol.~17, no.~10, pp. 1733--1746, Oct
  2015.

\bibitem{Mesaros2018_TASLP}
A.~Mesaros, T.~Heittola, E.~Benetos, P.~Foster, M.~Lagrange, T.~Virtanen, and
  M.~D. Plumbley, ``Detection and classification of acoustic scenes and events:
  Outcome of the {DCASE} 2016 challenge,'' \emph{IEEE/ACM Transactions on
  Audio, Speech, and Language Processing}, vol.~26, no.~2, pp. 379--393, Feb
  2018.

\bibitem{Mesaros2019_TASLP}
A.~Mesaros, A.~Diment, B.~Elizalde, T.~Heittola, E.~Vincent, B.~Raj, and
  T.~Virtanen, ``Sound event detection in the {DCASE} 2017 challenge,''
  \emph{IEEE/ACM Transactions on Audio, Speech, and Language Processing}, 2019,
  in press.

\bibitem{Adavanne2018_JSTSP}
\BIBentryALTinterwordspacing
S.~Adavanne, A.~Politis, J.~Nikunen, and T.~Virtanen, ``Sound event
  localization and detection of overlapping sources using convolutional
  recurrent neural networks,'' \emph{IEEE Journal of Selected Topics in Signal
  Processing}, vol.~13, no.~1, pp. 34--48, March 2018. [Online]. Available:
  \url{https://ieeexplore.ieee.org/abstract/document/8567942}
\BIBentrySTDinterwordspacing

\bibitem{Adavanne2019_DCASE}
\BIBentryALTinterwordspacing
S.~Adavanne, A.~Politis, and T.~Virtanen, ``A multi-room reverberant dataset
  for sound event localization and uetection,'' in \emph{Proceedings of the
  Workshop on Detection and Classification of Acoustic Scenes and Events
  (DCASE)}, 2019. [Online]. Available:
  \url{http://dcase.community/documents/challenge2019/technical\_reports/DCASE2019\_Adavanne.pdf}
\BIBentrySTDinterwordspacing

\bibitem{politis2020dataset}
\BIBentryALTinterwordspacing
A.~Politis, S.~Adavanne, and T.~Virtanen, ``A dataset of reverberant spatial
  sound scenes with moving sources for sound event localization and
  detection,'' \emph{arXiv e-prints: 2006.01919}, 2020. [Online]. Available:
  \url{https://arxiv.org/abs/2006.01919}
\BIBentrySTDinterwordspacing

\bibitem{politis2020overview}
\BIBentryALTinterwordspacing
A.~Politis, A.~Mesaros, S.~Adavanne, T.~Heittola, and T.~Virtanen, ``Overview
  and evaluation of sound event localization and detection in dcase 2019,''
  \emph{IEEE/ACM Transactions on Audio, Speech, and Language Processing},
  vol.~29, pp. 684--698, 2020. [Online]. Available:
  \url{https://https://arxiv.org/abs/2009.02792}
\BIBentrySTDinterwordspacing

\bibitem{Naranjo-Alcazar2020_task3_report}
J.~Naranjo-Alcazar, S.~Perez-Castanos, J.~Ferrandis, P.~Zuccarello, and
  M.~Cobos, ``Task 3 dcase 2020: Sound event localization and detection using
  residual squeeze-excitation cnns,'' DCASE2020 Challenge, Tech. Rep., July
  2020.

\bibitem{roy2018concurrent}
A.~G. Roy, N.~Navab, and C.~Wachinger, ``Concurrent spatial and channel
  ‘squeeze \& excitation’in fully convolutional networks,'' in
  \emph{International Conference on Medical Image Computing and
  Computer-Assisted Intervention}.\hskip 1em plus 0.5em minus 0.4em\relax
  Springer, 2018, pp. 421--429.

\bibitem{naranjoalcazar2020acoustic}
J.~{Naranjo-Alcazar}, S.~{Perez-Castanos}, P.~{Zuccarello}, and M.~{Cobos},
  ``Acoustic scene classification with squeeze-excitation residual networks,''
  \emph{IEEE Access}, vol.~8, pp. 112\,287--112\,296, 2020.

\bibitem{Shimada2021}
K.~Shimada, Y.~Koyama, N.~Takahashi, S.~Takahashi, and Y.~Mitsufuji, ``Accdoa:
  Activity-coupled cartesian direction of arrival representation for sound
  event localization and detection,'' in \emph{IEEE International Conference on
  Acoustics, Speech and Signal Processing (ICASSP)}, Toronto, Ontario, Canada,
  June 2021.

\bibitem{hu2018squeeze}
\BIBentryALTinterwordspacing
J.~Hu, L.~Shen, and G.~Sun, ``Squeeze-and-excitation networks,'' \emph{2018
  IEEE/CVF Conference on Computer Vision and Pattern Recognition}, Jun 2018.
  [Online]. Available: \url{http://dx.doi.org/10.1109/CVPR.2018.00745}
\BIBentrySTDinterwordspacing

\end{thebibliography}

%
% or list them by yourself
% \begin{thebibliography}{9}
% 
% \bibitem{dcase2016web}
%   \url{http://www.cs.tut.fi/sgn/arg/dcase2016/}.
%
% \bibitem{IEEEPDFSpec}
%   {PDF} specification for {IEEE} {X}plore$^{\textregistered}$,
%   \url{http://www.ieee.org/portal/cms_docs/pubs/confstandards/pdfs/IEEE-PDF-SpecV401.pdf}.
%
% \bibitem{PDFOpenSourceTools}
%   Creating high resolution {PDF} files for book production with 
%   open source tools, 
%   \url{http://www.grassbook.org/neteler/highres_pdf.html}.
%
% \bibitem{eWilliams1999}
% E. Williams, \emph{Fourier Acoustics: Sound Radiation and Nearfield Acoustic
%   Holography}. London, UK: Academic Press, 1999.
% 
% \bibitem{ieeecopyright}
%   \url{http://www.ieee.org/web/publications/rights/copyrightmain.html}.
%
% \bibitem{cJones2003}
% C. Jones, A. Smith, and E. Roberts, ``A sample paper in conference
%   proceedings,'' in \emph{Proc. IEEE ICASSP}, vol. II, 2003, pp. 803--806.
% 
% \bibitem{aSmith2000}
% A. Smith, C. Jones, and E. Roberts, ``A sample paper in journals,'' 
%   \emph{IEEE Trans. Signal Process.}, vol. 62, pp. 291--294, Jan. 2000.
% 
% \end{thebibliography}

\end{sloppy}
\end{document}